# Remote-contact catalysis for target-diameter semiconducting carbon nanotube array

Jiangtao Wang[1,‡,*], Xudong Zheng[1,‡], Gregory Pitner[2], Xiang Ji[1], Tianyi Zhang[1], Aijia Yao[1], Jiadi Zhu[1], Tomás Palacios[1], Lain-Jong Li[3], Han Wang[2], Jing Kong[1,*]

[1] Department of Electrical Engineering and Computer Science, Massachusetts Institute of Technology, Cambridge, Massachusetts, USA
[2] Taiwan Semiconductor Manufacturing Company, Corporate Research, San Jose, California, USA
[3] Taiwan Semiconductor Manufacturing Company, Corporate Research, Hsinchu, Taiwan
*Corresponding authors: wangjt@mit.edu (J.W.); jingkong@mit.edu (J.K.)
‡These authors contributed equally to this work.

## Abstract

Electrostatic catalysis has been an exciting development in chemical synthesis (beyond enzymes catalysis[1]) in recent years, boosting reaction rates and selectively producing certain reaction products[2]. Most of the studies to date have been focused on using external electric field (EEF) to rearrange the charge distribution in small molecule reactions such as Diels-Alder addition[3], carbene reaction[4], etc. However, in order for these EEFs to be effective, a field on the order of 1 V/nm (10 MV/cm) is required, and the direction of the EEF has to be aligned with the reaction axis[5]. Such a large and oriented EEF will be challenging for large-scale implementation, or materials growth with multiple reaction axis or steps. Here, we demonstrate that the energy band at the tip of an individual single-walled carbon nanotube[6] (SWCNT) can be spontaneously shifted in a high-permittivity growth environment, with its other end in contact with a low-work function electrode (e.g., hafnium carbide or titanium carbide[7]). By adjusting the Fermi level at a point where there is a substantial disparity in the density of states (DOS) between semiconducting (s-) and metallic (m-) SWCNTs[8], we achieve effective electrostatic catalysis for s-SWCNT growth assisted by a weak EEF perturbation (200V/cm). This approach enables the production of high-purity (99.92%) s-SWCNT horizontal arrays with narrow diameter distribution (0.95±0.04 nm), targeting the requirement of advanced SWCNT-based electronics for future computing[9-11]. These findings highlight the potential of electrostatic catalysis in precise materials growth, especially for s-SWCNTs, and pave the way for the development of advanced SWCNT-based electronics[12].

**Main text**

Electrostatic catalysis has garnered significant attention in recent years as a promising approach to improve the selectivity and accelerate the rates of chemical reactions by utilizing an EEF[2,5]. This has opened up new possibilities for designing and tailoring reactions with precision, including catalytic processes such as Diels-Alder addition[13] or Ullmann coupling reactions[14] at single-molecule level. However, despite this potential, practical implementation of electrostatic catalysis on a large scale has been hampered by the large EEF required, with extremely high voltages (~10 MV needed for 1 cm sample size). Furthermore, for the synthesis of materials with more complex structures either involving multiple reaction axis or multiple reaction steps, such oriented EEF will not be effective[5]. In this work, we expand the scope of electrostatic catalysis to the growth of one-dimensional materials with a weak EEF perturbation (200-V/cm square wave). We report an interesting phenomenon of remote-contact catalysis, occurring at the catalyst end of a single-walled carbon nanotube (SWCNT) during its synthesis when the opposite end is in contact with a low work-function metal. We found that in the synthesis environment where the permittivity is high, band bending along the SWCNT not only occurs normally at the contact interface with the metal contact, but also extends to the remote end that could be tens of microns away. Such remote band-bending effectively shifts the Fermi level of the SWCNT, leading to a spontaneous electrostatic energy separation between metallic (m-) and semiconducting (s-) SWCNT during their catalytic growth. As a result, only a weak perturbation from an EEF is needed to initiate a stable chirality twist[15] of an m-SWCNT to an s-SWCNT. The produced SWCNT horizontal arrays show a high semiconducting purity of 99.92% and a narrow diameter distribution (0.95±0.04 nm) which is exactly what the semiconductor industry has been looking for towards the high-performance and energy-efficiency semiconductor materials beyond silicon[9-12].

The development of high-performance, low-power chips is a critical area of focus in modern electronics, aiming to meet the ever-increasing demands for computing power within energy-efficient constraints. In the quest for materials that can bridge these goals, semiconducting single-walled carbon nanotubes (s-SWCNTs) stand out as a promising material because of their ballistic electron transport and superior electrostatic control[16,17]. As a typical one-dimensional quantum material, the bandgap of s-SWCNTs is highly sensitive to their diameters, which is crucial for tuning their electrical properties[18]. This sensitivity is captured by the equation $E_g = \frac{0.85\ [eV \cdot nm]}{d_{CNT}\ [nm]}$, highlighting how the bandgap ($E_g$) inversely proportional to the nanotube diameter ($d_{CNT}$). A bandgap of 0.85 eV, corresponding to a diameter of 1.0 nm, has been identified as optimal for contemporary technology, striking a balance between reducing leakage-current and preserving on-current performance[9,12,19]. Due to the diverse structure of SWCNTs and their very similar physical chemistry properties, the preparation of high-quality SWCNTs horizontal arrays with the optimal target diameter and high semiconducting purity has become a central challenge[9,20]. The SWCNT preparation techniques currently fall into two categories: direct growth and solution-based post-processing. The latter method is known for achieving very high semiconducting purity but faces challenges such as defects, large-area alignment, and a wide diameter distribution often deviating from the optimal 1nm target[21-24]. In contrast, while the direct growth approach excels in high-quality and large-area alignment, it has yet to match the semiconducting purity and precise diameter control seen in requirements of semiconductor industry[25-37]. Recognizing these challenges, in this work, we have found this unexpected phenomenon of remote-contact catalysis both new and intriguing but also shows its potential for selective growth of high-purity s-SWCNT horizontal arrays with the optimal target diameter.

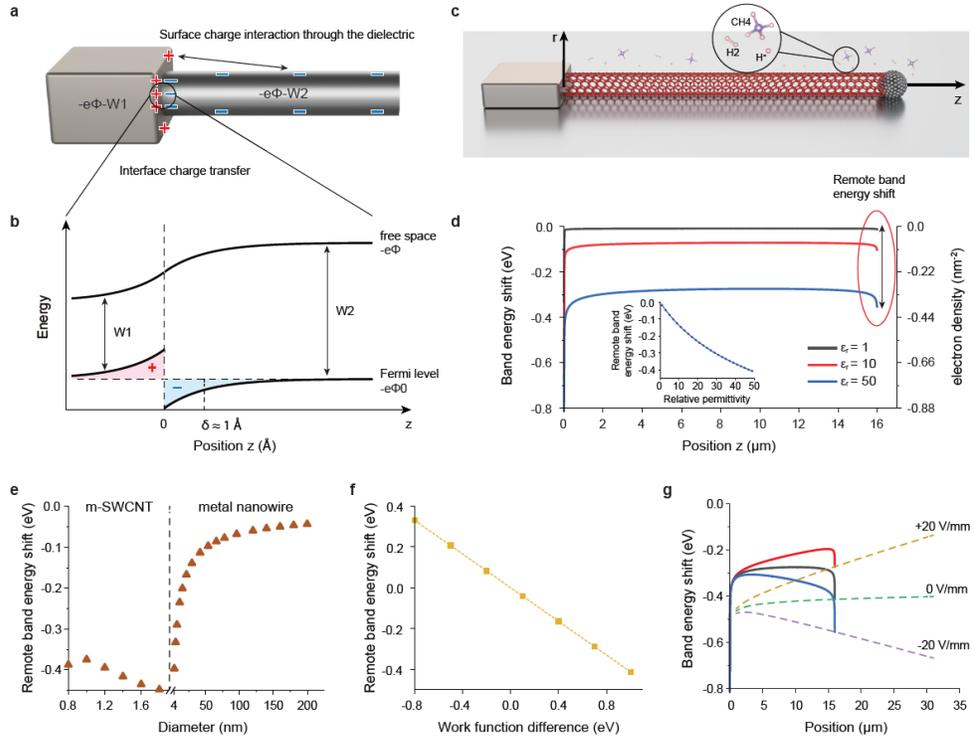

**Fig. 1 | Band energy shift in SWCNT in a high permittivity environment. a**. The illustration of the charge distribution between two metals with different work functions. A metal rod is on the right-hand side, contacted by a metal electrode on the left. $W_1$ and $W_2$ are the work function of the two metals. **b**. The theoretical band bending at the interface between the two metals. **c**. The illustration of an m-SWCNT contacted by a low-work function metal in a high-permittivity environment. **d**. The simulated band energy shift along the direction of the carbon nanotube (i.e., the z direction). Inset: The remote band energy shift (i.e. at the end of the SWCNT) of the m-SWCNT as a function of various relative permittivity. The work function difference (i.e., between $W_1$ and $W_2$) and length of the m-SWCNT are fixed as 1 eV and 16 μm, respectively. **e**. The remote band energy shift of an m-SWCNT or a metal (graphite) nanowire versus its diameter when contacted by a metal electrode. The relative permittivity, work function difference, and length of the m-SWCNT (or nanowire) are fixed as 50, 1 eV, and 16 μm, respectively. **f.** The remote band energy shift with various contact work function difference. The relative permittivity and length of the m-SWCNT are fixed as 50, 16 μm, respectively. The dotted line is a linear fitting to the simulated data. **g.** The simulated band energy shift under various EEF, respectively. The solid lines represent the band energy shift along a 16-μm

m-SWCNT. The dash lines correspond to the remote band energy shift values under various EEF as a function of the SWCNT length, thus the remote band energy shift at the nanotube tip end takes the minimal value with positive EEF, and has the maximum value with the negative EEF.

Band bending and charge transfer are well-known phenomena that occur at the interface of two materials with different work functions, such as between metals, between metals and semiconductors, or between metal nanoparticles and oxide substrates[38,39]. Typically, the range of interface charge transfer and band bending inside the conventional metals is limited to the immediate vicinity of the contact interface due to the screening effect[40] (Fig. 1a,b). The characteristic decay length of the band bending is about 0.1 nm (see Methods for derivation). However, for materials with nanometer dimensions, the surface charge plays a more important role. When considering the charge redistribution at the outer surface of the two metals in Fig. 1a, three factors should be taken into account: (1) the surface charge interaction through the environment ambient which acts as a dielectric, (2) the surface charge-band energy shift correlation, and (3) the surface charge conservation. To incorporate both electrostatics and band theory, we employ a global potential (which encompasses all potentials outside, inside, and at the surface of the metal, taking infinity as zero), denoted as $\phi$, to describe the system. When the geometry and boundary potentials are fixed, the surface charge density is proportional to the environment dielectric constant,

$$\sigma = -\varepsilon_r \varepsilon_0 \frac{\partial \phi}{\partial n} \tag{1}$$

in which $\varepsilon_r$ and $\varepsilon_0$ are the relative permittivity and vacuum permittivity, respectively, $-\frac{\partial \phi}{\partial n}$ refers to the electric field perpendicular to the surface. On the other hand, according to the band theory, for the $i$th metal, the surface charge density is proportional to the band energy shift $\Delta E_i$ (for a constant density of state at 0K approximation),

$$\sigma_i = e \cdot DOS_i \cdot \Delta E_i \tag{2}$$

$$\Delta E_i = -(e\phi + W_i - e\phi_{0i}) \quad (3)$$

where $e$ represents the elementary charge, $DOS_i$ and $W_i$ refer to the density of state per area and the work function of the $i$th metal, respectively, and $\phi_{0i}$ is the potential inside the metal respect to the ground, $-e\phi_{0i}$ is the Fermi level (or quasi-Fermi level when an external voltage is applied). The surface charge of conservation is expressed as,

$$\sum_i \oiint \sigma_i \, dS = 0 \quad (4)$$

and $i$ represents the $i$th metal in the system. Since the global potential $\phi$ is the only variant defined, the model is based on electrostatics and incorporates band theory. Equations (1)-(4) establish the system's boundary conditions.

To accurately simulate the charge distribution and band bending along the metal rod surface (e.g., an m-SWCNT or a metal nanowire), we employed COMSOL to establish an electrostatic model using cylindrical coordinates. The m-SWCNT (or metal nanowire) with a radius denoted as $r_0$ is positioned at $r = 0$ along the positive direction of z axis, while the electrode surface is located at $z = 0$ with the assumption that the inner potential is zero (Fig. 1c). We have found that when a one-dimensional (1D) SWCNT is in contact with a low work-function metal electrode (such as HfC, 3.5 eV[7]) in a high-permittivity environment, the band energy along the tube entirely shifts with an additional remote band bending at the other end of the SWCNT (Fig. 1d, red circled region). Our simulations revealed that the charge distribution and band energy shift $\Delta E$ along the m-SWCNT are strongly influenced by the relative permittivity $\varepsilon_r$ of the environment (Fig. 1d & inset). When $\varepsilon_r$ is close to 1, the screening effect from free carriers in the SWCNT dominates and the band bending regions are limited to the immediate vicinity of the contact interface. However, when $\varepsilon_r$ is more than ten, the band energy shift $\Delta E$ of the whole SWCNT becomes significant due to the strong charge interaction through the dielectric. The band bending at the remote end of the m-SWCNT (which will be referred to as the remote band energy shift, $\Delta E$ at the remote end) is formed because of the long aspect ratio. In the growth environment of the SWCNTs, the relative permittivity of the growth environment is typically close to 50 due to the presence of charged ions generated at high temperatures[41,42] (1073K) (see Method and Extended Data Fig. 1 for measurement results).

For a better understanding of the remote band energy shift, we compare the difference between a thick metal nanowire and an m-SWCNT. Unlike m-SWCNTs, metal nanowires in our simulations possess two end surfaces. Additionally, while the surface DOS of metal nanowire remains constant across different nanowire diameters, the surface DOS of m-SWCNT is inversely proportional to its diameter due to quantum confinement. Figure 1e shows the remote band energy shift in relation to the diameter for both an m-SWCNT and a metal nanowire (in an $\varepsilon_r = 50$ growth environment). The remote band energy shift for a 1D m-SWCNT is clearly noticeable (~0.4eV here). Considering charge conservation and the fact that the total energy states of a thicker nanowire is much more than that in an m-SWCNT, the surface charge density as well as the remote band energy shift of the bulk metal are substantially reduced. On the contrary, the remote band energy shift does not vary a lot with m-SWCNT diameter, because of the inverse relationship between its DOS and diameter. As a result, pronounced and stable remote band energy shifts should be primarily observed with SWCNTs in high-permittivity environments (note that an m-SWCNT is used here for simplicity of the DOS near its Fermi level, the result should be similar with an s-SWCNT during the synthesis process).

We have additionally observed that the remote band energy shift is proportional to the work function difference between the m-SWCNT and the contacted metal, and slightly decreases with increasing length of the m-SWCNT (Fig. 1f, and see Extended data Fig. 2 for more details). The band energy shift in the m-SWCNT is affected by applying a weak EEF, because of the high permittivity and the large aspect ratio of m-SWCNT. The solid lines in Fig. 1g show the band energy shift along the 16-μm m-SWCNT when +20 V/mm, 0 V/mm, and -20 V/mm EEFs are applied, respectively. The dash lines show the remote band energy shift with different m-SWCNT length. It is worth noting that the polarity and average value of the band energy shift are pinned by the remote contact although the EEF modulation becomes more effective when the m-SWCNT is longer.

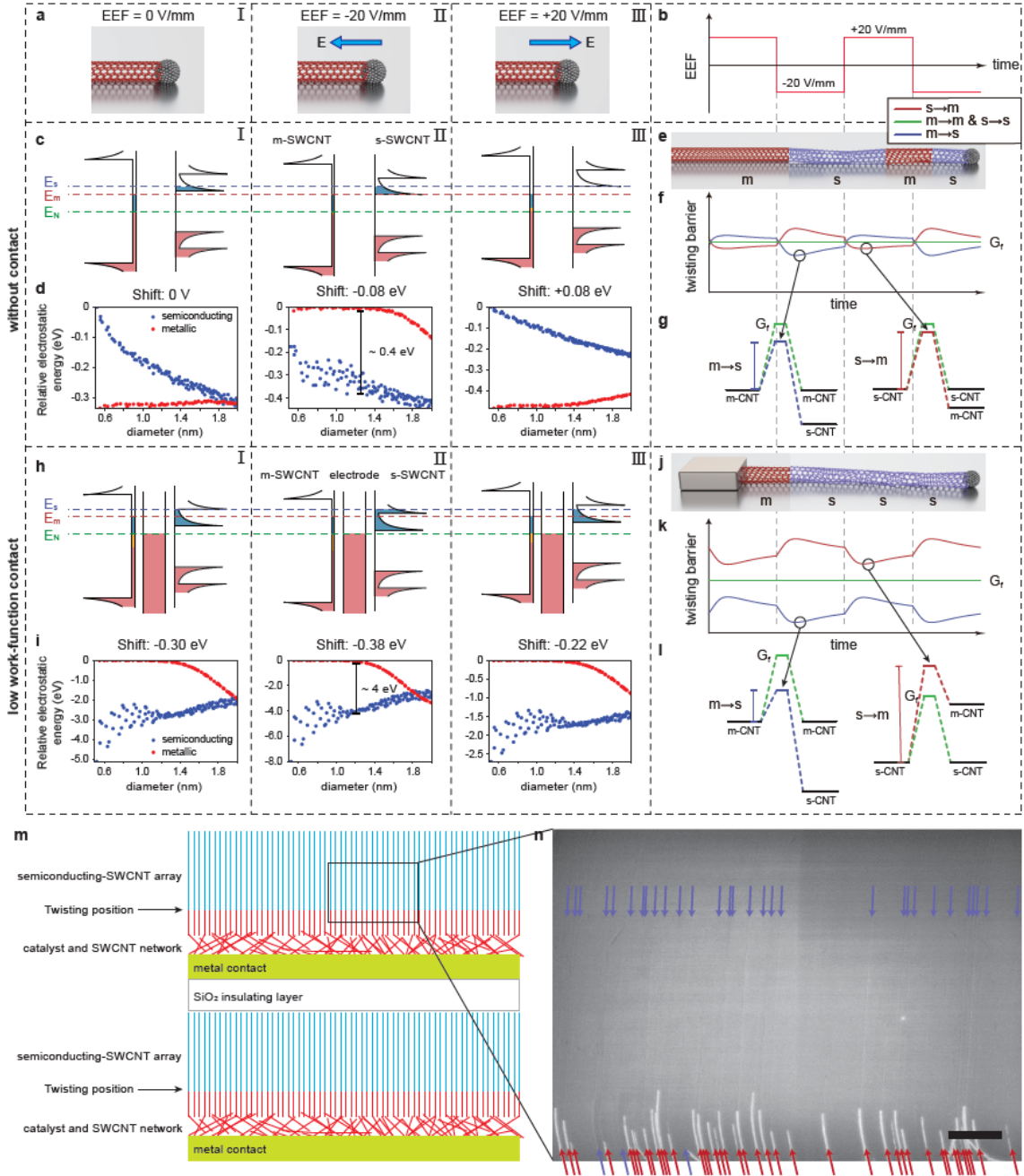

**Fig. 2 | Remote band bending as a catalyst for selective growth of s-SWCNTs. a,b.** The schematic of the direction and waveform of the applied EEF perturbation. **c,h.** The illustration of the band filling of s-SWCNT and m-SWCNT under different EEF without contact and with low work-function contact, respectively. $E_s$ and $E_m$ refer to the growth induced spontaneous charging level for s- and m-SWCNT, respectively. $E_N$ is the Fermi level of the tubes without growth. **d,i.** The calculated relative electrostatic energy of

SWCNTs under zero (panel I), -20V/mm (panel II), +20V/mm (panel III) EEF without (**d**) and with remote band energy shift (**i**) of -0.30eV. The zero point of relative electrostatic energy is defined as the maximum value within the data frame. **e,j**. The illustration of unstable chirality twist and stable chirality twist, respectively. **f,k**. The evolution of the twisting barrier of s→m, m→m/s→s, and m→s with time under an alternating EEF. "s" and "m" indicate s-SWCNT and m-SWCNT, respectively. **g,l**. The schematic of the twisting barrier variation causing unstable and stable twist, respectively. **m.** The illustration of the catalyst substrate layout that comprises (low work-function) metal contacts, catalyst strips, and $SiO_2$ insulating layers with the aligned SWCNT on the substrate (blue is the semiconducting region and red represent metallic dominant region). **n**, The SEM image of the stable twist of horizontally aligned SWCNTs from initially mixed metallic and semiconducting chiralities to only semiconducting chiralities. The red and blue arrows indicate m-SWCNTs and s-SWCNTs, respectively. Scale bar: 2 μm.

Such remote band energy shift turns out to be an effective electrostatic catalyst for the selective growth of s-SWCNTs. Fig. 2a & b illustrates a SWCNT with its tip end in a weak alternating square-wave EEF. It is worth noting that, in our system, the growth mode of the SWCNTs is tip growth[15]. Fig. 2c illustrates the DOS of an m-SWCNT and s-SWCNT with its Fermi level shift under such EEF. In a typical CNT growth, the catalyst/SWCNT interface is charged[41], and the large difference of quantum capacitance between m-SWCNT and s-SWCNT separates the electrostatic energy plot of SWCNTs into two branches[15] (Fig. 2d, panel I, which is calculated under the assumption that all SWCNTs grow with the same rate; see Methods for derivation). Compared with the high growth temperature (1073K), such a small energy difference between the m-SWCNTs and s-SWCNTs is not large enough to preferentially grow one type (either m- or s-) instead of another. Previously we have found that the negative half cycle of an alternating EEF perturbation can be used to prompt the m-SWCNTs to change into s-SWCNTs due to the lower electrostatic energy of s-SWCNTs under negative EEF[15] (as shown in panel II of Fig. 2d, under an EEF which causes a band energy shift of -0.08eV, the electrostatic energies of the s-SWCNTs becomes lower than m-SWCNTs). However, during the positive half cycle of the EEF, because the electrostatic energies of m-SWCNTs are lower (Fig. 2d panel III), the s-SWCNTs have a possibility to change back to m-SWCNTs. This is also illustrated in the twisting barrier change in Fig. 2e-g (see Methods for the derivation of twisting barriers). In contrast, when a low-work function metal is used to contact the SWCNT, the remote band energy shift effectively separates the electrostatic energies of s-SWCNTs and m-SWCNTs (Fig. 2h illustrates the DOS and energy level positions under 0, negative and positive EEFs while Fig. 2i panel (I-III) present the electrostatic energies of the s- and m-SWCNTs under the three scenarios). As a result, the twisting barriers are well separated, and each EEF switching will always prompt m-SWCNTs to be twisted into s-SWCNTs. Based on such understanding, we have carried out the growth of aligned SWCNTs on quartz substrates. Fig. 2m illustrates the layout of the growth substrates (single crystal quartz[43]) and the grown SWCNTs and Fig. 2n shows the scanning electron microscope (SEM) image of a corresponding growth result. It can be seen that all the SWCNTs have been twisted into s-SWCNT which shows darker contrast under the SEM imaging[44-46]. In fact, there are many ways to twist the chirality of the SWCNT during the synthesis, as long as an energy perturbation is input into the system (such as temperature change). Extended Data Fig. 3 shows the twisting results via temperature perturbation[47,48]. However, EEF perturbation shows much higher twisting efficiency (Extended Data Fig. 4) than temperature perturbation.

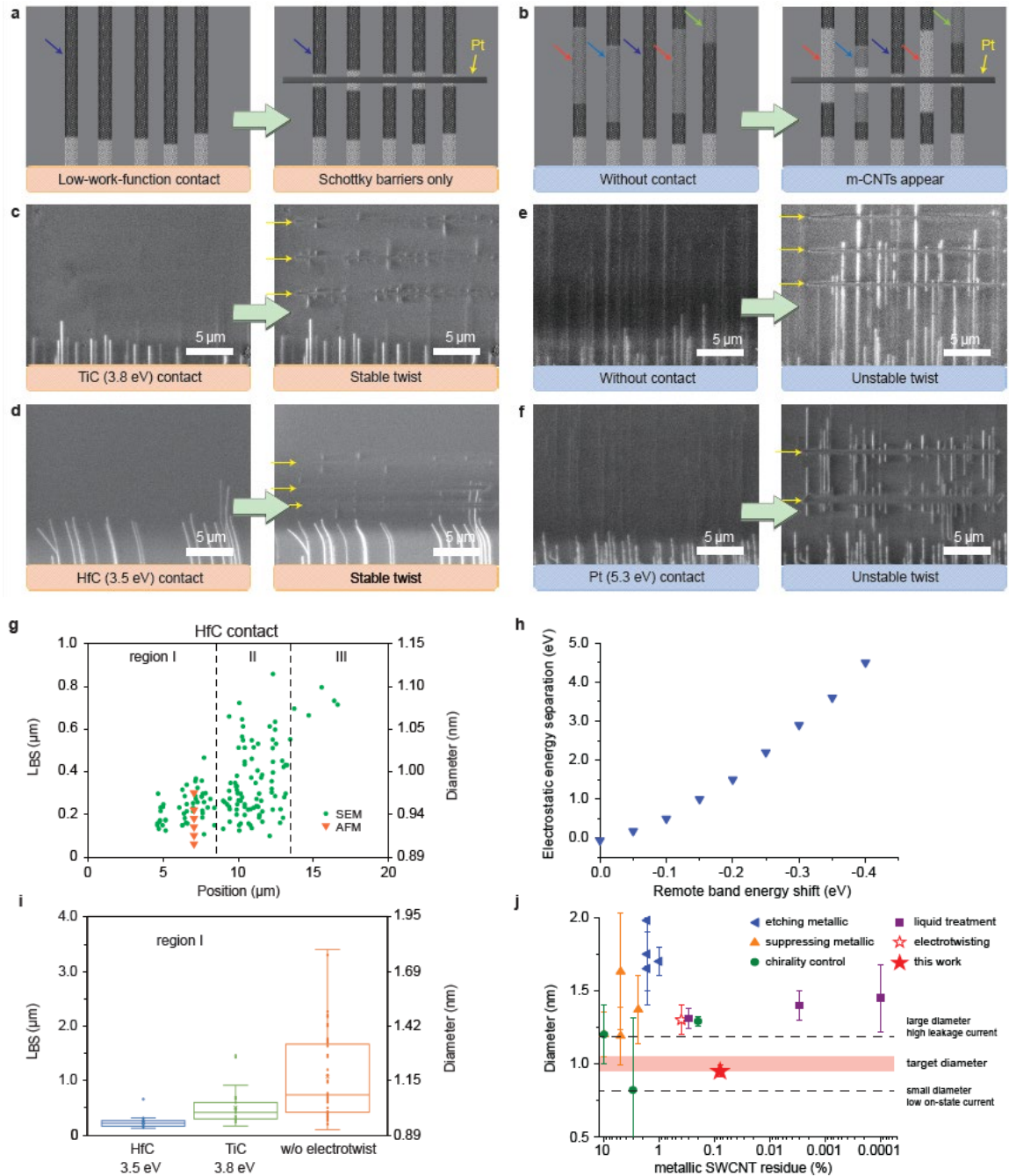

**Fig. 3 | The diametric evolution of the twisted SWCNTs under the effect of the remote band energy shift. a,b.** The schematic of identification of s-SWCNTs under SEM by contacting with a metal strip. Before contacting with the metal strip, bright nanotubes and dark nanotubes indicate the m-SWCNTs and uncertain SWCNTs before and after electrotwist, respectively. After contacting with the Pt strip, uncertain SWCNTs are identified by the charge transfer length shown as lightened segment in SEM images.

Dark blue and light blue arrows indicate the small-diameter s-SWCNTs and large-diameter s-SWCNTs, respectively. Red arrows refer to m-SWCNTs, and green arrows head to still uncertain parts. **c,d**. SEM identification of SWCNTs after stable electrotwist with TiC and HfC contact, respectively. **e,f**. SEM identification of SWCNTs after unstable twist without contact and with Pt contact, respectively. Scale bars: 5 μm. **g**. The diametric evolution of s-SWCNTs during the growth with HfC-contact under electrotwist. $L_{BS}$ refers to the length of the bright segment that is caused by the charge transfer when an s-SWCNT contacts with a Pt strip deposited *in-situ* under the SEM. Regions I-III are outlined based on the diameter distribution. AFM data (see Extended Data Fig. 6c&d) is also plotted for verification. **h**. The dependence of electrostatic energy separation between m- and s-SWCNTs on the remote band energy shift (here calculated for SWCNTs having diameters around 1.2 nm as an example). **i**. The comparison of diameter distributions of s-SWCNTs under HfC-contact electrotwist, TiC-contact electrotwist, and without electrotwist in region I. **j**. The state-of-the-art diameter control for both direct-growth methods and post-treatment methods (see Extended Data Table 1 for the references list). The error bars represent the standard deviation of the diameter distribution. The metallic SWCNT residue of this work is determined by electrical measurements.

In addition to the observation of remote catalysis for selective growth of s-SWCNTs, we also found interesting diametric evolution of the twisted SWCNTs under the effect of the remote band energy shift, which gives a very narrow distribution of the diameter when the remote band energy shift is strong. In order to measure the diameter of the SWCNTs effectively, we utilized SEM imaging technique based on previously developed method[44-46]. For SWCNTs on an insulating substrate, under the electron beam illumination in SEM, the substrate becomes positively charged. The contrast of SWCNTs under SEM is determined by whether or not electrons are being compensated by a metal contact. Bright contrast indicates that the SWCNT is metallic and electrons are being compensated by its contact to a metal. Conversely, dark contrast indicates that there is no electron compensation, either because the SWCNT is semiconducting or because the SWCNT is not connected to a metal (and itself is not long enough to provide for the compensation). In our SEM imaging, we utilized electron beam induced deposition of Pt strips orthogonal to the aligned SWCNTs to observe the dark-contrast s-SWCNT better and study the diameters. This is illustrated in Fig. 3a & b, the left panel shows the aligned SWCNTs that are grown, the bright segments are m-SWCNTs connected with the bottom low work-function contact or the SWCNT networks (Fig. 2m & Extended Data Fig. 5) before being twisted by the EEF, and the dark segments indicate uncertain SWCNTs after the twisting induced by EEF, in which the electrons cannot be compensated. After depositing the Pt strips across the uncertain SWCNTs, the electrons can transfer from the Pt strips to the uncertain SWCNTs within the length of bright segments ($L_{BS}$) (right panels of Fig. 3a & b). For the m-SWCNT/Pt contact, the electrons could transfer freely to the whole segment, resulting in a long $L_{BS}$. In contrast, for the s-SWCNT/Pt contact, the charge transfer was limited by the Schottky barrier, resulting in a relatively short $L_{BS}$. The length of Schottky barrier-limited charge transfer region $L_{BS}$ near the Pt strip is proportional to the diameter of an s-SWCNT[44]. We classified the contrast of uncertain SWCNTs and $L_{BS}$ into three groups. The first group had very dark contrast of SWCNTs with $L_{BS}$ shorter than 1.5 μm, indicating the s-SWCNTs with smaller diameter and larger bandgap (dark blue arrows in Fig. 3a & b). The second group had lighter dark contrast of SWCNTs with $L_{BS}$ between 1.5 μm to 3 μm, indicating the s-SWCNTs with larger diameter and smaller bandgap (pale blue arrows in Fig. 3b). The third group had bright contrast of SWCNTs with the $L_{BS}$ longer than 3 μm, usually assigned to m-SWCNT segments that were not directly connected with the bottom electrode or the SWCNT network (red arrows in Fig. 3b). The green arrows in Fig. 3b point to the still uncertain SWCNT segments with lighter dark contrast that do not directly contact with the Pt strips. Based on the above knowledge, we compared the effect of remote-contact catalysis with low-work function contact under the alternating EEF and the same growth of samples with no metal contact. Fig. 3c & d show the SEM images of s-SWCNTs grown with low-work function contact, in which TiC (3.8eV[7]) and HfC (3.5eV[7]) are used, respectively. After *in-situ* depositing the Pt strips, it was found that only short $L_{BS}$ appear at the contact points for both TiC and HfC cases, showing stable electrotwist. This means the m-SWCNTs were twisted to s-SWCNT, and the s-SWCNTs do not twist back to m-SWCNTs. In contrast, Fig. 3e shows when no metal contact is being used, it shows multiple s-SWCNTs were twisted back to m-SWCNTs, indicating obvious unstable electrotwist. These experimental observations are in good consistent with our theoretical understanding shown in Fig. 2 e-g and further verified the effects of the remote-contact catalysis. To compare with the effect of low work-function contact, we also used high work-function (Pt, 5.3eV[7]) as a metal contact and observed similar unstable twist, which again confirms our understanding of the remote-contact catalysis (see Extended Data Fig. 2a for calculations of high work-function contact).

With the $L_{BS}$ measurements along the growth direction of the SWCNTs, we studied the diametric evolution effect upon continuous electrotwist under the alternating EEF. Fig. 3g shows the scatter diagram of the Schottky barrier-limited charge transfer length $L_{BS}$ of s-SWCNTs (contacted with Pt strips) versus the growth position starting from the low work-function HfC contact edge. On the basis of the distribution of the data points, it was found that the diametric evolution can be divided into three regions. In region I, the growth position is closer than 8.5 μm, in which the diameters of s-SWCNTs are obviously thinner and have a narrower distribution. In region II, when the growth position is between 8.5 and 13.5 μm, the diameters of s-SWCNTs increase and have a broader distribution. In region III, the diameters of s-SWCNTs slightly increase further when the growth position is farther than 13.5 μm. This observation aligns with our calculations presented in Fig. 1g where the remote band energy shift exhibits more pronounced oscillations in response to the length of the SWCNT under a square-waveform EEF (this is plotted again in Extended Data Fig. 7a, with regions I-III labeled). The upper bound of the oscillation corresponds to the minimal remote band energy shift, the absolute value of which decreases with the length of the SWCNT as it grows. From Fig. 3h, the relative electrostatic energy separation between s-SWCNTs and m-SWCNTs decreases as the absolute value of the remote band energy shift decreases, therefore, as the SWCNT grows longer, the relative electrostatic energy separation also decreases. Extended Data Fig. 7b-d shows the scatter diagrams of the relative electrostatic energy vs. the diameter of SWCNTs with three upper bound values (red dots in Extended Data Fig. 7a) of the remote band energy shift in these three regions. From all these results it can be seen that s-SWCNTs with smaller diameter have lower electrostatic energy, but the slope of decreasing (within a diameter range of 0.9-1.1 nm) is steeper in region I than in region II, and region III is the shallowest. As a result, the growth of s-SWCNTs with smaller diameter would be favorable, in which the arrow indicated the direction of the diametric evolution of s-SWCNTs in region I. This tendency is also balanced by the energy cost in terms of the elastic energy of the very thin SWCNT wall, which will limit the decrease of the diameter of s-SWCNTs. According to our observation, the diameters are concentrated around 0.95 nm (see Fig. 3g) for region I. Such a comparison helps to understand the diameter broadening in region II and III, nevertheless, the overall diameters of the s-SWCNTs are less than 1.2nm in these regions. In Fig. 3c & d, we can see the average length of $L_{BS}$ at s-SWCNTs/Pt contact in the case of HfC is shorter than that in the case of TiC, which means the diameters of s-SWCNTs using HfC contact are smaller than those using TiC contact. Figure 3i shows the detailed comparison of diameter distributions with HfC-contact electrotwist, TiC-contact electrotwist, and without electrotwist, based on the corresponding SEM images in region I (see Method and Extended Data Fig. 6 for details). It is evident that lower work function of the contact delivers a narrower distribution of s-SWCNT diameter, this is consistent with our calculations in Extended Data Fig. 7. In the condition of using HfC contact, the diameter distribution of as-synthesized s-SWCNTs is 0.95±0.04 nm, which shows a great potential in the diameter control. For example, it can be anticipated that with an *in-situ* SWCNT length monitoring during the growth process[33], one only needs to apply the EEF perturbation when the SWCNTs are growing in region I, and stop the EEF perturbation as the SWCNTs grow to longer distances, in this way, arrays of s-SWCNTs with 0.95±0.04 nm diameters can be obtained. Furthermore, since the elastic energy within the SWCNT wall decreases at higher temperatures, the energy equilibrium point for the diameter can be adjusted by varying the growth temperature[48]. Here our theoretical analysis and calculations are not only consistent with the experimental observation, but also revealed the great potential of remote-contact catalysis for the precise control of the structure of SWCNTs. Figure 3j shows the state-of-the-art diameter control for both direct growth and post-treatment methods[15,22-32,49-54]. Recent studies of band-to-band tunneling leakage in CNT transistors claim ~1.0 nm diameter is optimal for high-speed energy-efficient logic devices, reinforcing utility of this approach for semiconductor applications[55].

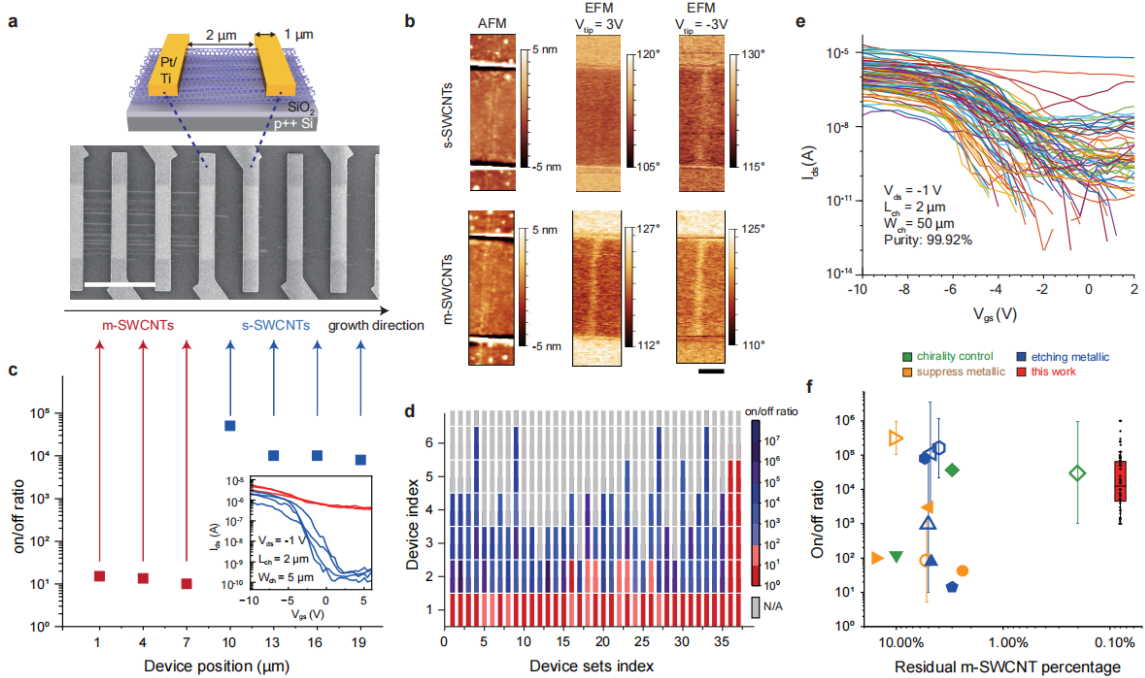

**Fig. 4 | The electrical assessment of the s-SWCNT arrays grown with remote-contact catalysis. a.** Device schematic and basic structural parameters of the s-SWCNT field-effect transistors (FET), and an SEM image of a device set containing multiple FETs fabricated along the SWCNT growth direction. Scale bar: 5 μm. **b.** The AFM and corresponding EFM images of an s-SWCNT and an m-SWCNT in devices. Scale bar: 500 nm. **c.** The evolution of measured on/off ratio along the growth direction of one SWCNT array. Inset: the transfer characteristic curves of each channel. **d.** The mapping of on/off ratio of as-grown SWCNT arrays. Channel width: 50 μm. **e.** All the measured transfer characteristic curves for extracting the purity of s-SWCNT. **f.** The benchmark of as-grown s-SWCNTs among various CVD growth methods. The error bars indicate a range of on/off ratio reported by the corresponding references. (see Extended Data Table 1 for the references list)

To further verify the effectiveness of the electrotwist with remote-contact catalysis, we fabricated arrays of field effect transistors (FETs) with as-synthesized SWCNTs and electrically measured the semiconducting purity. The schematic and SEM images in Fig. 4a show one set of the as-fabricated back-gated FET devices based on one SWCNT array (Fig. 2m, see Methods for details). Multiple contact electrodes are placed in an array along the growth direction to form a device set (each array contains 4-8 devices depending on the length of the SWCNTs) so that the SWCNTs before and after electrotwist position are electrically characterized. Figure 4b shows the phase image of electrostatic force microscopy (EFM) measurements on SWCNTs inside the device channel, with the height image of atomic force microscopy (AFM) being shown as a reference. When the tip bias ($V_{tip}$) is changed from 3V to -3V, a clear contrast change from dark to bright is observed on s-SWCNTs. In contrast, the m-SWCNTs before electrotwist remains bright despite what $V_{tip}$ is applied. The corresponding transfer characteristic curves ($I_{ds}$-$V_{gs}$) are plotted in Fig. 4c inset and the extracted on/off ratio are plotted in Fig. 4c as a function of the device position. Along the growth direction, the first three SWCNT devices are conductive with on/off ratios ~ 10 (plotted in red), indicating the presence of m-SWCNTs at the early growth stage before the electrotwist. However, the fourth device shows p-type semiconducting behavior with on/off ratio ~$10^5$ and the rest of devices all exhibit on/off ratios ~$10^4$ (plotted in blue). It is evident that all the SWCNTs in the channel becomes semiconducting after electrotwist with remote-contact catalysis and stays semiconducting stably without being twisted back to metallic, which agrees with the change of SWCNTs contrast in the SEM image (Fig. 4a). The on/off ratio of the s-SWCNTs, which is inversely proportional to their diameter, also showed a consistent evolution with the diametric evolution as shown in Fig. 3g. Extended Data Figure 8 shows the large-area electrical measurements over the whole sample, which presents similar results as in Fig. 4c. Most of the devices changes from metallic into semiconducting with on/off ratio >$10^3$ at device position between 7 μm and 10 μm. The slight difference in twist position is attributed to the difference in growth speed at different substrate locations.

To accurately estimate the electrotwisting yield and the semiconducting purity of produced SWCNT array, we repeated the electrical measurements over many SWCNT arrays with a much wider channel width (50 μm, to include more SWCNTs in one channel). Figure 4d shows the on/off ratio mapping of 37 such device sets in one growth batch (Extended Data Fig. 9a-d). For each device set index, similar FET measurements in Fig. 4c are repeated. Once the SWCNTs are twisted into semiconducting, they stay semiconducting and do not twist back to metallic under weak EEF perturbation (Extended Data Fig. 9e). There are only two untwisted device sets at the substrate edge on this chip, a careful examination under SEM reveals three m-SWCNTs running through these two device sets (Extended Data Fig. 9f, g). Figure 4e shows the transfer curves of all the SWCNT devices where electrotwist was supposed to happen. Two transfer curves from these two untwisted device sets were taken respectively and were also shown in Fig. 4e. Based on all the electrical assessments, the actual semiconducting purity of synthesized SWCNTs was extracted to include ~3740 SWCNTs and estimated to be ~99.92%, which further proved both high efficiency and high stability of electrotwist with the remote-contact catalysis. Figure 4f compares the semiconducting purity and on/off ratio in this work to the previous reports for high-density SWCNTs[25-32,49-54] (see Extended Data Table 1 for reference details). The residual m-SWCNT percentage in this work is record-low for CVD-based enrichment methods while a high on/off ratio is maintained. Furthermore, the purity approaches the level of polymer sorted semiconducting CNTs which suffer from residue and alignment challenges when used in electronics[10].

It is worth noting that the origin of remaining three untwisted m-SWCNTs is possibly due to their earlier nucleation and fast growth. Extended Data Fig. 10a shows the SEM image of a growth result intentionally stopped before applying the weak EEF perturbation. It can be seen that one m-SWCNT grew much longer than the other tubes. Comparing with Extended Data Fig. 10b that shows the growth result with normal electrotwist applied, we found that the density and length of long m-SWCNT in these two images are both very similar. It was very possible that the early-grown m-SWCNT stopped its growth before the formation of SWCNT network and connection with the low-work function metal, or had already grown very long (Extended Data Fig. 10c shows a specific example) before the electrotwist. In other words, the growth of long m-SWCNTs were not within the control window of electrotwist. These observations suggest that the current semiconducting purity is limited by the early growth rather than the electrotwist and remote-contact catalysis. A future effort to improve the semiconducting purity even further should focus on ensuring all the grown SWCNTs to establish electrical contact to the electrode at early stage of their growth. It is anticipated that in the near future, the concept of remote-contact catalysis will not only be crucial for the synthesis of high-purity s-SWCNT arrays, but will also help to advance electrostatic catalysis in the precise synthesis of other low-dimensional materials, including control of band structure, ferroelectricity, and heterojunctions.

**Data availability**

All data needed to evaluate the conclusions herein are present in the Article.

**Code availability**

The python codes used in relative electrostatic energy calculation can be available at doi:xxxxxxxxx.

**Acknowledgement:** This work was sponsored in part by TSMC Corporate Research, J.W. and J. K. also acknowledge the support from the U.S. Army Research Office (ARO) MURI project under grant number W911NF-18-1-04320431. X. Z. and J. K. acknowledge the support by the US Army Research Office grant number W911NF2210023. T.Z. and J.K. acknowledge the support by the U.S. Department of Energy (DOE), Office of Science, Basic Energy Sciences (BES) under award DE-SC0020042. Part of this work was performed at Stanford Nanofabrication Facility, supported by the National Science Foundation under award ECCS-2026822.